\documentclass[twocolumn,prb,superscriptaddress,floatfix]{revtex4-1} 

\usepackage[dvipdfmx]{graphicx}
\usepackage{amsmath,amssymb,bm}
\usepackage{xcolor}
\usepackage{float}
\usepackage{ulem}

\usepackage[
colorlinks=true,
citecolor=blue,
setpagesize=false]{hyperref}

\begin{document}

\title{
Proximate Kitaev system for an intermediate magnetic phase in in-plane magnetic fields
}

\author {Beom Hyun Kim}
\affiliation{Korea Institute for Advanced Study, Seoul 02455, South Korea}
\author {Shigetoshi Sota}
\affiliation{
  Computational Materials Science Research Team, 
  RIKEN Center for Computational Science (R-CCS),  
  Kobe, Hyogo 650-0047, Japan}
\author {Tomonori Shirakawa}
\affiliation{
  Computational Materials Science Research Team, 
  RIKEN Center for Computational Science (R-CCS),  
  Kobe, Hyogo 650-0047, Japan}
\author {Seiji Yunoki}
\affiliation{
  Computational Materials Science Research Team, 
  RIKEN Center for Computational Science (R-CCS),  
  Kobe, Hyogo 650-0047, Japan}
\affiliation{
  Computational Condensed Matter Physics Laboratory, 
  RIKEN Cluster for Pioneering Research (CPR), 
  Saitama 351-0198, Japan}
\affiliation{Computational Quantum Matter Research Team, RIKEN, 
  Center for Emergent Matter Science (CEMS), Wako, Saitama 351-0198, Japan}
\author{Young-Woo Son}
\affiliation{Korea Institute for Advanced Study, Seoul 02455, South Korea}

\date{\today} 
\begin{abstract}
Motivated by the magnetic phase transition of a proximate Kitaev system 
$\alpha$-RuCl$_3$ in the presence of a magnetic field,
we study the simplest but essential quantum spin model with the ferromagnetic nearest neighboring (NN) 
Kitaev interaction and additional antiferromagnetic third NN Heisenberg interaction.
Employing both exact diagonalization and density matrix renormalization group methods, we demonstrate that
the model shows the magnetic phase transition from the zigzag order phase 
to the spin polarized phase through an intermediate phase in both cases 
when an in-plane magnetic field is applied perpendicular to the NN bond direction 
and when an out-of-plane field is applied,
in good agreement with experimental observations.
Furthermore, we verify that additional symmetric off-diagonal $\Gamma$ 
interaction and ferromagnetic Heisenberg interaction between NN spins
can both suppress the intermediate phase with the in-plane field.
Our result gives important clues on determining relevant interactions
in the field-induced magnetic phase transition of proximate Kiteav systems.
\end{abstract}

\maketitle

{\it Introduction} --
Quantum spin liquid (QSL) is an exotic quantum phase 
in which any magnetic long-range order is prevented due to
strong quantum fluctuation~\cite{Savary2016}.
The Kitaev model with directional Ising-type interactions 
between the nearest neighboring (NN) spins (Kitaev interaction) 
in a honeycomb lattice
is an exactly solvable system to host as the ground state the QSL phase
interpreted with free Majorana fermions in a static $Z_2$ gauge field~\cite{Kitaev2006}.
For the last decade, there have been considerable efforts devoted to search materials
hosting the Kitaev interaction~\cite{Winter2017a,Takagi2019,Motome2020}.

Among those candidates, $4d$/$5d$-based honeycomb systems 
such as $\alpha$-RuCl$_3$ and Na$_2$IrO$_3$
were proposed as best ones to have strong Kitaev interactions
\cite{Jackeli2009,HSKim2015}.
While such systems certainly possess the predominant Kitaev interaction,
their magnetic ground state has been turned out to be not
the Kitaev spin liquid (KSL) but an antiferromagnetic (AFM) phase with the zigzag order 
\cite{Singh2010,Sears2015}.
Other non-negligible types of magnetic interactions 
such as Heisenberg interactions and symmetric off-diagonal $\Gamma$ interactions [see Eq.~(\ref{Hamil})]
have been known to play a role in determining the non-Kitaev ground state~\cite{Rau2014,Winter2016}.

Despite its long-range magnetic order, 
$\alpha$-RuCl$_3$ has been thought as a proximate Kitaev system~\cite{BHKim2016}.
Recent experiments with inelastic neutron scattering and Raman spectroscopy 
evidenced magnetic continuum excitations attributed to
possible fractionalized Majorana fermions~\cite{Sandilands2015,Banerjee2016,Banerjee2017,Do2017}. 
A fractionalized magnetic entropy has been also observed 
in specific heat measurements~\cite{Do2017,Widmann2019}.
Moreover, a lot of experimental studies have supported 
that the zigzag order can be suppressed and 
an intermediate phase (IP), possible QSL, can emerge 
in between the zigzag spin order and the spin polarized order when an external magnetic field is applied
\cite{Kubota2015,Sears2017,Leahy2017,Baek2017,Wang2017a,Zheng2017,Wolter2017,
Banerjee2018,Kasahara2018,Jansa2018,Lampen-Kelley2018a,Wellm2018,Balz2019,Yokoi2020}.
This IP has been recently reported to appear in the in-plane field
(especially $a$-axis field) as well as the out-of-plane field
\cite{Lampen-Kelley2018a,Balz2019,Yokoi2020}.
An observed half-integer quantized plateau in 
thermal Hall conductivity has highly promoted that an anticipated IP would be
the KSL~\cite{Kasahara2018,Yokoi2020}.
The nature and origin of the IP are still under debate.

\begin{figure*}[!tb]
\centering
\includegraphics[width=1.8\columnwidth]{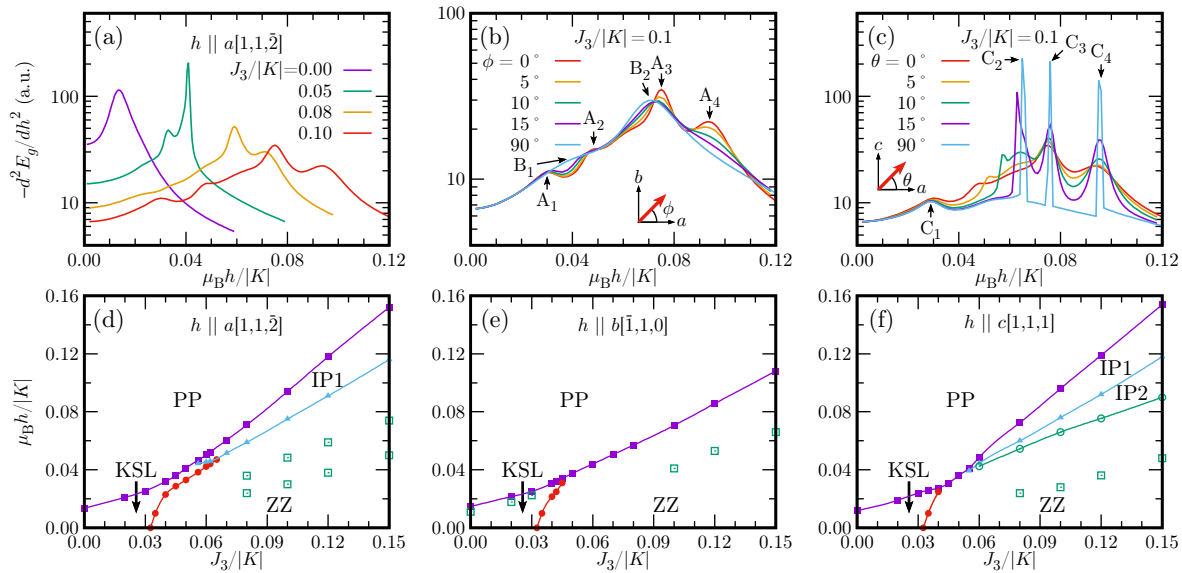}
\caption {
  (a) Second derivative of the ground state energy ($d^2E_g/dh^2$) 
  with respect to the magnetic field $h$ along the $a$ axis ([1,1,$\bar{2}$] direction~\cite{supp})
  for various values of the third nearest neighboring interaction $J_3$.  
  (b)--(c) Second derivative of the ground state energy
  for various field directions (b) in the $ab$ plane 
  and (c) in the $ac$ plane when $J_3=0.1|K|$. 
  Here, $\phi$ and $\theta$ refer to angles between the $a$ axis and the field direction 
  in the $ab$ and $ac$ planes, respectively.
  (d)--(f) Ground state magnetic phase diagrams as a function of 
  the external magnetic field $h$ along (d) the $a$ axis,
  (e) the $b$ axis ([$\bar{1}$,1,0] direction~\cite{supp}), and 
  (f) the $c$ axis ([1,1,1] direction~\cite{supp}).
  `A$_{1-4}$' `B$_{1,2}$', and `C$_{1-4}$' 
  in (b) and (c) represent peaks of $-d^2E_g/dh^2$ in the $a$-, $b$-, and
  $c$-axis fields, respectively.
 `KSL', `ZZ', `PP', and `IP1 (IP2)' in (d)--(f) refer to the Kitaev spin liquid, 
  zigzag order, polarized phase, and intermediate phase, respectively. 
  Green squares indicate peak positions corresponding to peaks 
  A$_1$ and A$_2$ in (d), B$_1$ in (e), and C$_1$ in (f).
 All results are obtained for the $K$-$J_3$ model with $K<0$
 calculated on a periodic $24$-site cluster using the ED method~\cite{supp}. 
}
\label{fig_kj3}
\end{figure*}

The effective model for $\alpha$-RuCl$_3$ 
in the presence of an external magnetic field has been proposed
with the following Hamiltonian: 
\begin{align}
H &= \sum_{\gamma \langle i,j \rangle_\gamma} 
 \left[ J \mathbf{S}_i \cdot \mathbf{S}_j + K S_{i\gamma}S_{j\gamma} 
+ \Gamma \left( S_{i\alpha}S_{j\beta} 
+ S_{i\beta}S_{j\alpha} \right) \right] 
\nonumber \\
&+ \sum_{\gamma \langle i,j \rangle_\gamma}
\Gamma' \left( S_{i\alpha}S_{j\gamma} + S_{i\gamma}S_{j\alpha}
 + S_{i\beta}S_{j\gamma} + S_{i\gamma}S_{j\beta}
\right) \nonumber \\
&+ \sum_{ \langle\langle\langle i, j \rangle \rangle \rangle} 
J_3 \mathbf{S}_i \cdot \mathbf{S}_j
- \mu_B \sum_i \vec{h} \cdot 
\mathbf{g} \cdot \vec{S}_i ,
\label{Hamil}
\end{align}
where $\mathbf{S}_i$ is the spin-1/2 operator at site $i$ with its $\gamma$ ($=x,y,z$) component $S_{i\gamma}$, 
$\langle i, j \rangle_\gamma$ stands for the NN pair
of sites $i$ and $j$ along the $\gamma$ bond, and 
$\alpha$ and $\beta$ refer to the two remaining coordinates other than $\gamma$ 
[see Fig.~S1(b) in Supplemental Material (SM)~\cite{supp}]. 
$K$ and  $J$ are Kitaev and Heisenberg interactions, respectively, 
$\Gamma$ and $\Gamma'$ are two types of symmetric off-diagonal interactions, 
and $J_3$ is Heisenberg interaction between the 3rd NN sites. 
$\vec{h}$ is the external magnetic field, $\mathbf{g}$ is the $g$ tensor, and $\mu_B$ is the Bohr magneton. 
For simplicity, we assume an isotropic $g$ tensor, although
both fairly isotropic and highly anisotropic 
ones have been proposed before~\cite{Agrestini2017,Yadav2016}.

Until now, various models, in which specific parameters are set to zero
in the general model given in Eq.~(\ref{Hamil}), 
such as $J$-$K$~\cite{Banerjee2017,Jiang2019}, $K$-$\Gamma$~\cite{Ran2017,
Wang2017b,Gohlke2018,Catuneanu2018}, 
$K$-$\Gamma$-$\Gamma'$~\cite{Gordon2019,HYLee2020},
and $J$-$K$-$\Gamma$-$J_3$ models~\cite{Winter2017b,Winter2018}, 
have been adopted to explore the magnetic properties of $\alpha$-RuCl$_3$.
Extended models including additional interactions 
have also been introduced in some literatures
\cite{Yadav2016,Hou2017,Janssen2017,Eichstaedt2019,Laurell2020,Maksimov2020}.
Among them, some are based on the AFM Kitaev interaction ($K>0$).
In this case, an intermediate $U(1)$ QSL phase is stabilized 
regardless of the direction of the field~\cite{Hickey2019}.
The $K$-$J$ model exhibits the zigzag order when $K>0$ and $J<0$, 
and a clear IP, putative $U(1)$ QSL, can appear 
in the external field along the $c$ axis ([1,1,1] direction in terms of
local coordinates of spins depicted in Fig. S1(a)~in SM~\cite{supp})~\cite{Jiang2019}.
However, recent consensus is that the ferromagnetic (FM) Kitaev interaction ($K<0$) is
more likely in $\alpha$-RuCl$_3$~\cite{Winter2016,Do2017,Winter2017b,
Banerjee2018,Yadav2016,Sears2020}.
The models with $K<0$ have also been employed to explain the zigzag order and
the IP in the external field.
The $K$-$\Gamma$-$\Gamma'$ model with $K<0$, $\Gamma>0$, and $\Gamma'<0$ studied recently 
successfully gives the IP in the field along the $c$ axis.
However, the IP is totally missing when the field
is along the $a$ axis ([1,1,$\bar{2}$] direction~\cite{supp})~\cite{Gordon2019,HYLee2020}.
Numerical calculations for the $J$-$K$-$\Gamma$-$J_3$ model with $K<0$ also failed to 
show the IP in the $a$-axis field~\cite{Winter2018}.

In this study, we propose a simple theoretical quantum model only with $K$ ($<0$) and $J_3$ ($>0$)
to exhibit a genuine IP in both in-plane and out-of-plane magnetic fields.
With the help of exact diagonalization (ED) 
and density matrix renormalization group (DMRG) methods,
we demonstrate that the IP evidently manifests itself
in the field along both $a$ and $c$ axes,
whereas it collapses in the field along the $b$ axis ([$\bar{1}$,1,0] direction~\cite{supp}), 
in accordance with recent experiments~\cite{Jansa2018,Balz2019}.
Compared with other models to explain the ground state with the zigzag order,
we assert the important role of the 3rd NN interaction 
in the most promising Kitaev material $\alpha$-RuCl$_3$.

{\it Phase diagram of $K$-$J_3$ model} --
According to the ED calculation with a $C_3$ rotationally symmetric
24-site cluster [see SM~\cite{supp}], 
the $K$-$J_3$ model exhibits the phase transition from the KSL to the zigzag order phase
at $J_3/|K|\approx 0.033$.
When the external magnetic field is applied, the KSL or zigzag order phase
is suppressed and
the spin polarized phase is eventually stabilized in the strong field limit.
Figure~\ref{fig_kj3}(a) shows the second derivative of the ground state
energy, $d^2E_g(h)/dh^2$, with respect to the $a$-axis field for the KSL ($J_3/|K|=0$) 
and the zigzag order ($J_3/|K| = 0.05$, $0.08$, and $0.1$) phases.
Note that $-d^2E_g(h)/dh^2$ is proportional to the magnetic susceptibility.
Our calculation clearly shows the emergence of the IP
during the phase transition from the zigzag order to the polarized phase 
whereas the KSL phase transforms directly to the polarized phase
without any IP.

We further investigate the magnetic phase transition with various field directions.
Figures~\ref{fig_kj3}(b) and \ref{fig_kj3}(c) show
$-d^2E_g(h)/dh^2$ for $J_3/|K|=0.1$ 
when the field direction varies from the $a$ to $b$ axis in the $ab$ plane
and from the $a$ to $c$ axis in the $ac$ plane, respectively.
Interestingly, peak A$_4$ associated with the highest critical field at 
$\mu_Bh=0.094|K|$ gradually diminishes
when the field direction is away from the $a$ to $b$ axis.
This peak becomes indistinguishable when $\phi=15^\circ$,
where $\phi$ is an angle between the field direction and the $a$ axis in the $ab$ plane.
Only peak B$_2$ at $\mu_Bh=0.071|K|$ remains 
with a small satellite (peak B$_1$) at $\mu_Bh=0.041|K|$ in the $b$-axis field.
In terms of the symmetry of the $K$-$J_3$ model, the IP appears (disappears) 
when the field is applied perpendicular (parallel) to
one of three possible NN bonds,
in good accordance with recent experiments of $\alpha$-RuCl$_3$~\cite{Lampen-Kelley2018a,Yokoi2020}.

In contrast, when the field direction varies from the $a$ to $c$ axis in Fig.~\ref{fig_kj3}(c), 
the two highest peaks (peaks A$_3$ and A$_4$) are robust with 
even reducing their peak widths
until the field points close to the $[1,1,\bar{1}]$ direction, which 
corresponds to an angle $\theta$ between the field direction and the $a$ axis 
in the $ac$ plane being about 19.47$^\circ$ ($\approx \cos^{-1}\frac{2\sqrt{2}}{3}$).
Concomitantly, a new peak (peak C$_2$) appears around $\mu_Bh=0.066|K|$.
This directly indicates the emergence of additional IP 
when the field is applied out of the $ab$ plane.
Because the $K$-$J_3$ model is invariant under the $z$-component reversal of 
all spins ($S_{iz}\rightarrow -S_{iz}$),
$-d^2E_g(h)/dh^2$ is exactly the same in both magnetic fields along 
the $[1,1,\bar{1}]$ direction and the $[1,1,1]$ direction ($c$ axis).
Therefore, the IPs determined by peaks A$_3$ and A$_4$ in the $a$-axis field,
and peaks C$_3$ and C$_4$ in the $c$-axis field are adiabatically equivalent.

We should remark that peaks labeled as A$_1$ (at $\mu_Bh \approx 0.03|K|$), 
A$_2$ ($0.048|K|$), B$_1$ ($0.041|K|$), and C$_1$ ($0.028|K|$) in Figs.~\ref{fig_kj3}(b) and \ref{fig_kj3}(c), 
appearing in a small field region where the zigzag order still remains, 
are not related to the phase transition but due to the finite size effect [see SM~\cite{supp}].
We also confirm that the emergence of IP in both $a$- and $c$-axis fields
is robust in the ED calculation with a different cluster geometry,  
although the number of IPs in the $c$-axis field may depend on the cluster geometry 
[see SM~\cite{supp}].

Figures~\ref{fig_kj3}(d)--\ref{fig_kj3}(f) show the ground state phase diagrams of 
the $K$-$J_3$ model for three different field directions.
In these phase diagrams, the following features are noticeable:
i) The intermediate KSL phase is extended from $J_3/|K|\approx0.033$ 
to $0.065$ in the presence of the $a$-axis field 
but it progressively shrinks in other field directions.
ii) The IPs, emerging in both $a$- and $c$-axis fields,
are distinct from the KSL phase,
while any IP is inhibited in the $b$-axis field.
iii) In the $c$-axis field, two consecutive IPs 
occur
and the one (IP1) appearing in the larger field is adiabatically equivalent to the IP found 
in the presence of the $a$-axis field.

\begin{figure}[t]
\centering
\includegraphics[width=.95\columnwidth]{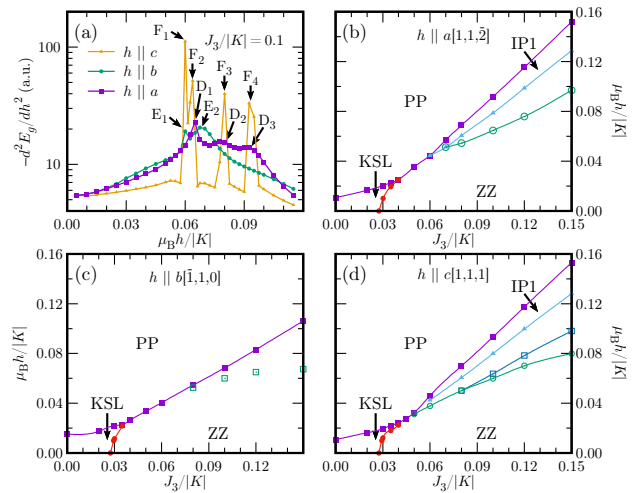}
\caption {
 (a) Second derivative of the ground state energy ($d^2E_g/dh^2$)
 with respect to the $a$-, $b$-, and $c$-axis fields $h$ 
 when $J_3/|K|=0.1$. 
 (b)--(d) Ground state magnetic phase diagrams 
  as a function of the external magnetic field $h$ along 
  (b) the $a$ axis, (c) the $b$ axis, and (c) the $c$ axis. 
  `D$_{1-3}$', `E$_{1,2}$', and `F$_{1-4}$' 
  in (a) represent peaks of $-d^2E_g/dh^2$ in the $a$-, $b$-, and
  $c$-axis fields, respectively. 
  Green squares indicate peak positions corresponding to peak E$_1$ in (c). 
  All results are obtained for the $K$-$J_3$ model with $K<0$
  calculated on a periodic $32$-site cluster using the DMRG method~\cite{supp}. 
}
\label{fig_DMRG}
\end{figure}

{\it DMRG calculation} --
For the robustness of our ED results shown above, especially to check the finite size effect,
we perform the DMRG calculation with a periodic 32-site cluster. 
The DMRG method is believed to accurately simulate 
the ground state of the $K$-$J_3$ model even in the two dimensional limit
[see SM~\cite{supp}].

Figure~\ref{fig_DMRG}(a) shows $-d^2E_g/dh^2$ 
for $J_3/|K|=0.1$ in the $a$-, $b$-, and $c$-axis fields.
Noteworthy, there is no peak of $-d^2E_g/dh^2$ below $\mu_B h/|K|=0.05$
in any field direction.
With this, we can assure that the low-field peaks below $\mu_B h/|K|=0.05$
in the ED calculation are due to the finite size effect 
and do not involve any phase transition.
Above $\mu_B h/|K|=0.05$,
the DMRG calculation finds three (D$_1$, D$_2$, and D$_3$), 
two (E$_1$ and E$_2$), and four peaks (F$_1$, F$_2$, F$_3$, and F$_4$) 
of $-d^2E_g/dh^2$ in the $a$-, $b$-, and $c$-axis fields, respectively,
in contrast with the ED calculation in which two, one, and three peaks appear, respectively.
Two and three types of consecutive IPs 
seem to be manifested in the $a$- and $c$-axis fields, respectively.
In the $b$-axis field, there exist two peaks (peaks E$_1$ and E$_2$).
The lower peak (peak E$_1$), however, is not related to the phase transition
since it extends into the zigzag phase in larger $J_3/|K|$ cases.
As in the ED calculation, the higher peak (peak E$_2$) determines the phase
boundary between the zigzag order and polarized phases.
Positions of highest two peaks (peaks D$_2$ and D$_3$) in the $a$-axis field 
and those of highest two peaks (peaks F$_3$ and F$_4$) in the $c$-axis field
are almost coincident with each other.
Therefore, our DMRG calculation further supports the adiabatic equivalence
between the two IPs (indicated as IP1 in Fig.~\ref{fig_DMRG}) 
appearing in the presence of the fields along the two directions.

The phase diagrams obtained by the DMRG calculation is summarized 
in Figs.~\ref{fig_DMRG}(b)--\ref{fig_DMRG}(d) for three different field directions.
Overall shapes are similar to those obtained by the ED calculation.
As shown in Fig.~\ref{fig_DMRG}(b), however, the intermediate KSL phase
does not exist for $0.04 \lessapprox \mu_B h \lessapprox 0.06$
in the $a$-axis field, different from the result of the ED calculation shown in Fig.~\ref{fig_kj3}(d).
We suspect that such a difference would result from the finite size effect. 
Moreover, it is not clear how many consecutive IPs exist in the $a$- and $c$-axis fields
because the 32-site calculation also suffers from the finite size effect.
Nevertheless, our study unambiguously reveals that
the IP is evident both in the $a$- and $c$-axis fields 
but it is absent in the $b$-axis field.

\begin{figure}[t]
\centering
\includegraphics[width=.95\columnwidth]{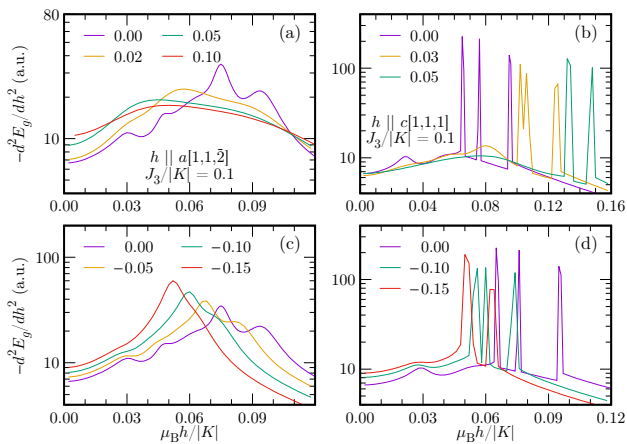}
\caption {
 Second derivative of the ground state energy ($d^2E_g/dh^2$) 
 with respect to the magnetic field $h$ 
 along (a) the $a$ axis for $\Gamma/|K|=0$, $0.02$, $0.05$, and $0.1$, 
 (b) the $c$ axis for $\Gamma/|K|=0$, $0.03$, and $0.05$, 
 (c) the $a$ axis for $J/|K|=0$, $-0.05$, $-0.1$, and $-0.15$, and 
 (d) the $c$ axis for $J/|K|=0$, $-0.1$, and $-0.15$,
 when $J_3/|K|=0.1$.
 All results are obtained for the $K$-$J_3$ model with $K<0$
 calculated on a periodic $24$-site cluster using the ED method~\cite{supp}. 
}
\label{fig_GM}
\end{figure}


{\it Discussion} --
As described in Eq.~(\ref{Hamil}), the relevant form of the magnetic interaction 
in $\alpha$-RuCl$_3$ has been believed to be more complex than the $K$-$J_3$ model.
The AFM $\Gamma$ interaction has been estimated to be comparable with
the $K$ interaction~\cite{Ran2017,Winter2017b,Sears2020,Janssen2017,Chaloupka2016,Wu2018,Lampen-Kelley2018b,Ozel2019}.
The AFM $J_3$ interaction has also been proposed to be as large as
the FM $J$ interaction with its strength being a few times weaker than 
$K$ interaction~\cite{Winter2016,Winter2017b,Maksimov2020}, 
or to be much weaker than the NN interactions~\cite{Wang2017b}.
Therefore, we clarify the role of the $\Gamma$ and $J$ terms in the field-induced phase transition of the $K$-$J_3$ model.
As shown in Fig.~\ref{fig_GM}(a) [Fig.~\ref{fig_GM}(c)], 
the multiple-peak structure in the $a$-axis field is almost suppressed and
a broad (sharp) one-peak structure remains even with a quite small value of 
$\Gamma$ ($J$) as large as $\Gamma/|K|\approx 0.05$ 
($J/|K|\approx -0.15$) for $J_3=0.1|K|$.
In the case of the $c$-axis field, the first IP determined by peaks C$_2$ and C$_3$ 
collapses when $\Gamma/|K| \approx 0.04$ ($J/|K| \approx -0.15$).
The second IP determined by peaks C$_3$ and C$_4$ is robust in 
the finite $\Gamma$ ($J$) interaction, 
although its boundaries shift upward (downward) with the overall IP region shrunk, 
as shown in Fig.~\ref{fig_GM}(b) [Fig.~\ref{fig_GM}(d)]. 
This clearly demonstrates why previous models
failed to give the IP in the $a$-axis field
in contrast with the case in the $c$-axis field~\cite{Gordon2019,HYLee2020,Winter2018}.

As shown in Fig.~\ref{fig_kj3}(d), the IP is clearly distinct from 
the KSL phase. Moreover, our model hardly identifies the nematic phase proposed
as the possible IP of the $K$-$\Gamma$-$\Gamma'$ model~\cite{HYLee2020}. 
We further verify that the magnetic phase of the classical $K$-$J_3$ model
continuously varies from the zigzag order to the fully polarized phase without any IP
regardless of the field direction~[see Ref.~\onlinecite{Janssen2017} and SM~\cite{supp}].
Thus, the classical magnetic phase can be ruled out for the IP.

Similar to the $K$-$J$ model with AFM $K$ and FM $J$ interactions, 
where $U(1)$ spin liquid was proposed for the IP~\cite{Jiang2019},
physical quantities such as the spin correlation function and the magnetization 
in our model calculations
show step-like variations around the phase boundaries in the $c$-axis field [see SM~\cite{supp}].
In the field away from the $c$ axis, the step-like variation is suppressed (not shown here). 
Such a feature remains in the $a$-axis field but entirely diminishes in the $b$-axis field.
Moreover, the magnetic excitation gap is almost closed at both $\Gamma$
and $M$ points, and the excitation spectrum is well-spread 
like continuum excitations at the $\Gamma$ point in the vicinity of the IP 
labeled as `IP1' in Figs.~\ref{fig_kj3}(d) and \ref{fig_kj3}(f) [see SM~\cite{supp}].
This likely implies that the IP1 is possibly $U(1)$ spin liquid 
characterized as the gapless fermionic excitation.
However, the entanglement entropy calculation of 
a periodic $2\times 16 \times 3$ cluster by the DMRG method 
reveals that the gap is only closed at the phase boundaries and 
the IP is gapped [see SM~\cite{supp}].
Therefore, our DMRG calculation indicates that 
the gapless $U(1)$ spin liquid may not be probable as the IP of the $K$-$J_3$ model, 
although the boundaries between the phases are still gapless.
Further studies are highly desired to identify the genuine nature of the IP.

{\it Conclusion} --
Based on the numerical calculation with both ED and DMRG methods,
we have found that the quantum $K$-$J_3$ model can exhibit the magnetic phase transition
from the zigzag order phase to the polarized phase via an IP even in the $a$-axis field. 
Our results are in good qualitative agreement with recent observations 
in the proximate Kitaev material $\alpha$-RuCl$_3$~\cite{Lampen-Kelley2018a,Balz2019,Yokoi2020}.
We also found that the IP in the $a$-axis field is adiabatically equivalent to the IP
in the $c$-axis field 
but diminishes in the $b$-axis field.
Considering that our model is quite concise,
our results provide a new insight on
determining a spin Hamiltonian relevant to $\alpha$-RuCl$_3$ and also 
on understanding its field-induced phase transition.

\begin{acknowledgments}
{\it Acknowledgments} --
We acknowledge Bongjae Kim, Seung-Hwan Do, Sungdae Ji, Kwang-Yong Choi,
Yong-Baek Kim, Hae-Young Kee, Eun-Gook Moon,
David A. S. Kaib, and Roser Valent\'i for fruitful discussion.
B.H.K. was supported by KIAS Individual Grants (CG068701). 
S.S. was supported by Grant-in-Aid for Young Scientists (B) (No.~JP17K14148) from MEXT, Japan. 
S.S and T.S. were supported by JST PRESTO (No.~JPMJPR191B and No.~JPMJPR191C), Japan. 
S.Y. was supported by Grant-in-Aid for Scientific Research (B) (No.~JP18H01183) from MEXT, Japan. 
Y.-W.S. was supported by KIAS Individual Grants (CG031509)
and by the NRF of Korea (Grant No. 2017R1A5A1014862, SRC Program: vdWMRC Center).
Numerical computations have been performed with 
the Center for Advanced Computation Linux Cluster System at KIAS
and the RIKEN supercomputer system (HOKUSAI GreatWave).
\end{acknowledgments}


\renewcommand{\thetable}{S\arabic{table}} 
\renewcommand{\thefigure}{S\arabic{figure}}
\renewcommand{\theequation}{S\arabic{equation}}
\renewcommand{\thesection}{S\arabic{section}}
\setcounter{table}{0}
\setcounter{figure}{0}
\setcounter{equation}{0}
\setcounter{section}{0}
\setcounter{enumi}{0} 
\renewcommand{\bibnumfmt}[1]{[S#1]}
\renewcommand{\citenumfont}[1]{S#1}

\onecolumngrid

\clearpage

\begin{center}
{\bf \Large
\textit{Supplemental Material}:\\
Proximate Kitaev system for an intermediate magnetic phase in in-plane magnetic fields
}

\vspace{0.2 cm}

Beom Hyun Kim,$^{1}$ Shigetoshi Sota,$^{2}$ 
Tomonori Shirakawa,$^{2}$   and Seiji Yunoki,$^{2,3,4}$ and Young-Woo Son$^{1}$

\vspace{0.1 cm}

{\small
{\it
$^1$Korea Institute for Advanced Study, Seoul 02455, South Korea

$^2$Computational Materials Science Research Team, 
  RIKEN Center for Computational Science (R-CCS), Kobe, Hyogo 650-0047, Japan

$^3$Computational Condensed Matter Physics Laboratory, 
  RIKEN Cluster for Pioneering Research (CPR), Saitama 351-0198, Japan

$^4$Computational Quantum Matter Research Team, RIKEN, 
  Center for Emergent Matter Science (CEMS), Wako, Saitama 351-0198, Japan
}
}

\end{center}

\section{ED calculation}

\begin{figure}[h!]
\centering
\includegraphics[width=.7\columnwidth]{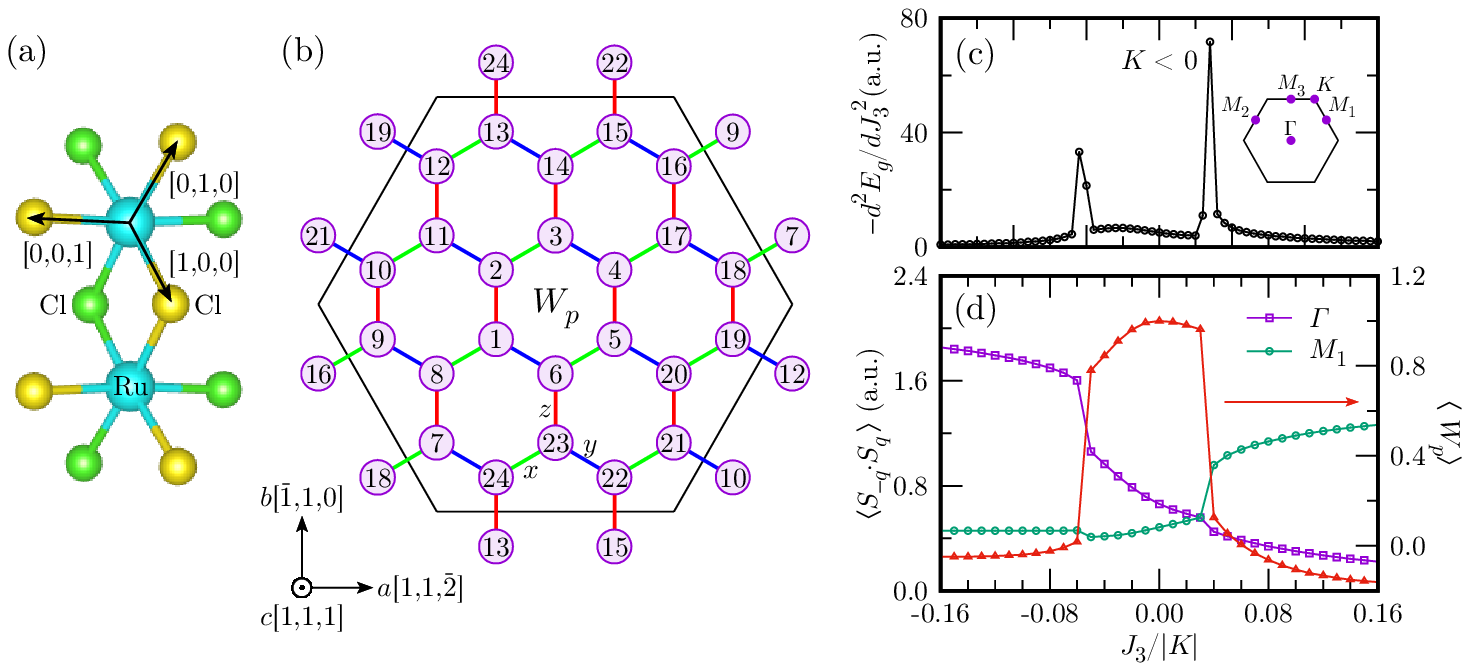}
\caption {
  (a) Schematic diagram of two Ru ions mediated via edge-sharing Cl ions
  along the $b$ axis in $\alpha$-RuCl$_3$. Cyan balls refer to Ru ions, and
  yellow and green balls represent Cl ions above and below the $ab$ plane, respectively.
  Black arrows denote local coordinate axes [1,0,0], [0,1,0], and [0,0,1]
  pointing from the central Ru ion to the upper Cl ions in a RuCl$_6$ octahedron, 
  which represent the local coordinates of spin $\mathbf{S}_i=(S_{ix},S_{iy},S_{iz})$.
  (b) Schematic diagram of a periodic 24-site cluster.
  Green, blue, and red lines refer to
  $x$-, $y$-, and $z$-type neighboring bonds, respectively. 
  Six sites in the central hexagon are used to calculate the expectation value of
  plaquette operator ($W_p$). 
  Lattice coordinates ($a$, $b$, and $b$ axes) are also indicated 
  as [1,1,$\bar{2}$], [$\bar{1}$,1,0], and [1,1,1] 
  in terms of the local coordinates of spins.
  (c) Second derivative of the ground state energy ($-d^2E_g/dJ_3^2$) 
  with respect to the third nearest neighboring interaction $J_3$ 
  and (d) static spin correlation functions $\left< \mathbf{S}_\mathbf{-q}\cdot\mathbf{S}_\mathbf{q}\right>$ 
  at the $\Gamma$ and $M_1$ points, 
  and the expectation value
  of the plaquette operator $W_p$ as a function of $J_3$ 
  for the $K$-$J_3$ model with the ferromagnetic Kitaev interaction ($K<0$) calculated on the periodic 24-site cluster 
  using the ED method.
  Brillouin zone of a honeycomb lattice is also shown in the inset of (c).
}
\label{fig_hc}
\end{figure}

To explore the ground state magnetic phase of the $K$-$J_3$ model,
we adopt a periodic 24-site cluster shown in Fig.~\ref{fig_hc}(b),
which is invariant under the $C_3$ rotation along the $c$ axis. 
Using the exact diagonalization (ED) method based on the Lanczos algorithm, we calculate the ground state and its energy.
In the pure Kitaev model ($J_3=0$), the ground state is the 
Kitaev spin liquid (KSL) and the expectation value of the plaquette operator 
$W_p=2^6 S_{1x}S_{2y}S_{3z}S_{4x}S_{5y}S_{6z}$ 
for one hexagon [see Fig.~\ref{fig_hc}(b)] should be exactly one.
The static spin correlation (SC) function 
$\left< \mathbf{S}_\mathbf{-q}\cdot\mathbf{S}_\mathbf{q}\right>$ 
does not show any peak structure at specific momentum $\mathbf{q}$ 
due to the quantum paramagnetism. 
Here, $\mathbf{S}_\mathbf{q}=\frac{1}{\sqrt{N}}\sum_{i=1}^N{\rm e}^{-i\mathbf{q}\cdot\mathbf{r}_i}\mathbf{S}_i$,  
$\mathbf{r}_i$ is the spatial location of the $i$th spin $\mathbf{S}_i$ in the honeycomb lattice [see Fig.~\ref{fig_hc}(b)], 
and $N$ is the number of sites. 
When $J_3$ is turned on, the ground state changes from the pure KSL phase.
Until the critical value of $J_3$, however, 
it is still adiabatically connected to the KSL phase
even though $\left< W_p \right>$ slightly deviates from one.
When the magnitude of $J_3$ is larger than the critical value ($J_3/|K|\approx 0.033$), 
$\left< W_p \right>$ is suddenly dropped and slowly saturated down to
a negative value ($\approx-0.19$) when $J_3/|K|>0$.
Concomitantly, 
$\left< \mathbf{S}_\mathbf{-q}\cdot\mathbf{S}_\mathbf{q}\right>$ at the 
$M$ points abruptly jumps.
This supports that the magnetic ground state changes from the KSL phase to
the zigzag order phase [see Fig.~\ref{fig_hc}(c) and \ref{fig_hc}(d)].

\section{Analysis of the zigzag order phase in the low-field limit}

\begin{figure}[h!]
\centering
\includegraphics[width=.9\columnwidth]{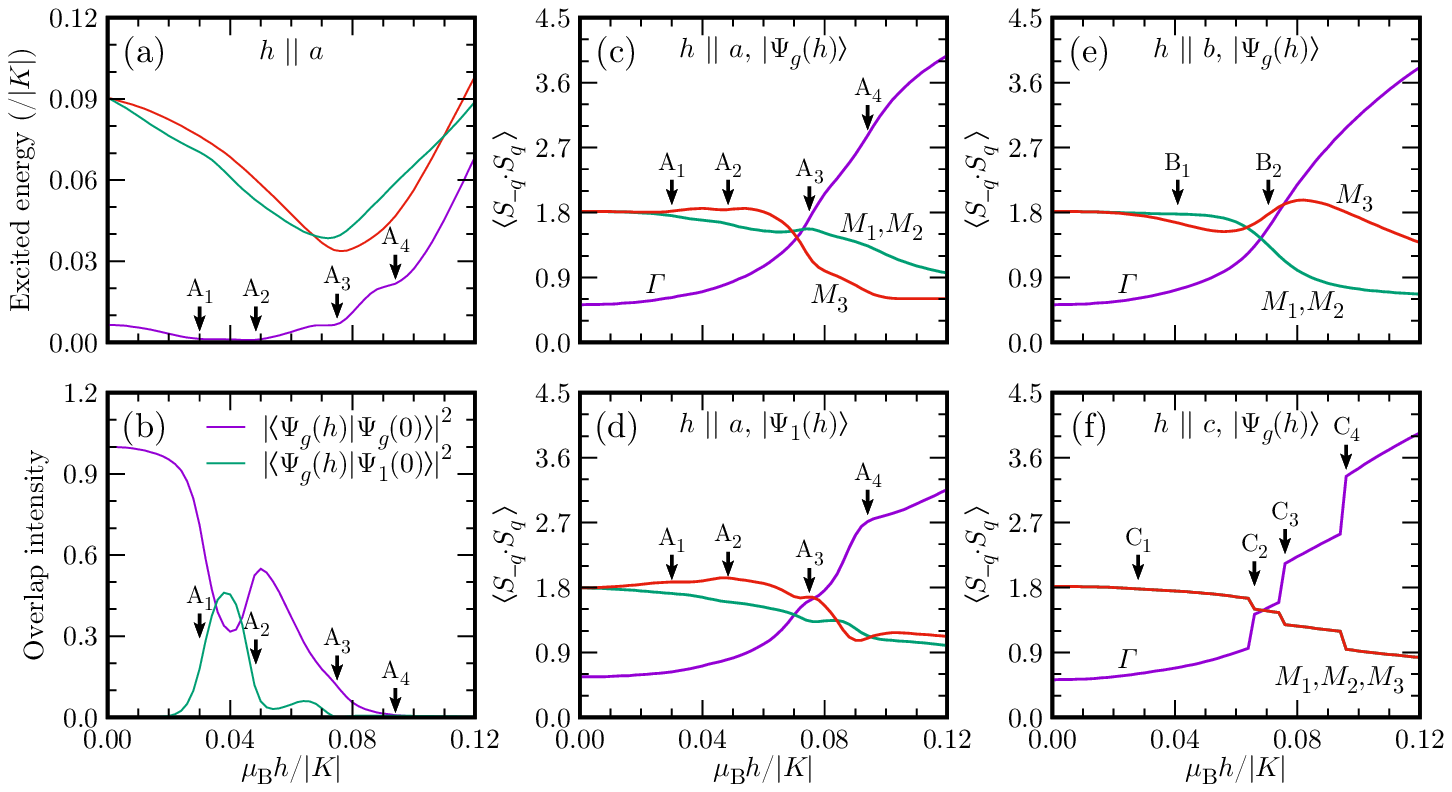}
\caption {
 (a) Evolution of the three lowest excitation energies with the $a$-axis field $h$.
 (b) The overlap intensity among the ground and first excited states,  
 $\left| \Psi_g(h)\right>$ and $\left| \Psi_1(h)\right>$, respectively,  
 in the magnetic field $h$ along the $a$ axis.  
 (c)--(f) Static spin correlation function 
 $\left< \mathbf{S}_\mathbf{-q}\cdot\mathbf{S}_\mathbf{q}\right>$
 at the $M_1$, $M_2$, $M_3$, and $\Gamma$ points
 for the ground states $\left| \Psi_g(h)\right>$ in the $a$-, $b$- and $c$-axis fields, 
 and the first excited state $\left| \Psi_1(h)\right>$ in the $a$-axis field. 
 Arrows indicate the positions of peaks A$_{1-4}$, B$_{1,2}$, 
 and C$_{1-4}$ of $-d^2E_g/dh^2$ [see Figs.~1(b) and 1(c) in the main text]. 
 All results are obtained for the $K$-$J_3$ model with the ferromagnetic Kitaev interaction ($K<0$) 
 and $J_3/|K|=0.1$ calculated 
 on a periodic 24-site cluster using the ED method. 
}
\label{fig_es}
\end{figure}

To explore the zigzag order phase more carefully, 
we examine the evolution of three lowest excited states with the $a$-axis field 
at $J_3/|K|=0.1$.
As shown in Fig.~\ref{fig_es}(a), the excitation energy of the first excited state is much smaller
than those of the second and third excited states in the low-field limit. 
When the magnetic field is within the region bounded by peaks A$_1$ and A$_2$ of $-d^2E_g/dh^2$ 
at $\mu_Bh/|K|\approx0.03$ and $0.05$, respectively [also see Fig.~1(b) in the main text], 
the first excitation energy becomes almost zero.
Also, $|\langle\Psi_g(h)|\Psi_g(0)\rangle|^2$ and $|\langle\Psi_g(h)|\Psi_1(0)\rangle|^2$ 
show the hollow and hump, respectively, in this region [see Fig.~\ref{fig_es}(b)].
It implies that the ground and first excited states in the zero field are certainly mixed together in this region, 
thus leading to the abrupt change of the ground and first excited states at the fields 
corresponding to peaks A$_1$ and A$_2$.
Note that the ground and first excited states both show dominant intensity 
of the static SC function $\left< \mathbf{S}_\mathbf{-q}\cdot\mathbf{S}_\mathbf{q}\right>$ at the $M$ points 
up to the magnetic field corresponding to peak A$_3$
[see Fig.~\ref{fig_es}(c) and \ref{fig_es}(d)]. 
Therefore, it hardly involves any genuine phase transition in this region.
Similarly, from the results shown in Fig.~\ref{fig_es}(e) and \ref{fig_es}(f), we can conclude that 
peaks B$_1$ and C$_1$ of $-d^2E_g/dh^2$ in the $b$- and $c$-axis fields 
[see Figs.~1(b) and 1(c) in the main text] are 
not related to a phase transition, either.

\section{Cluster geometry dependence}

\begin{figure}[h!]
\centering
\includegraphics[width=.9\columnwidth]{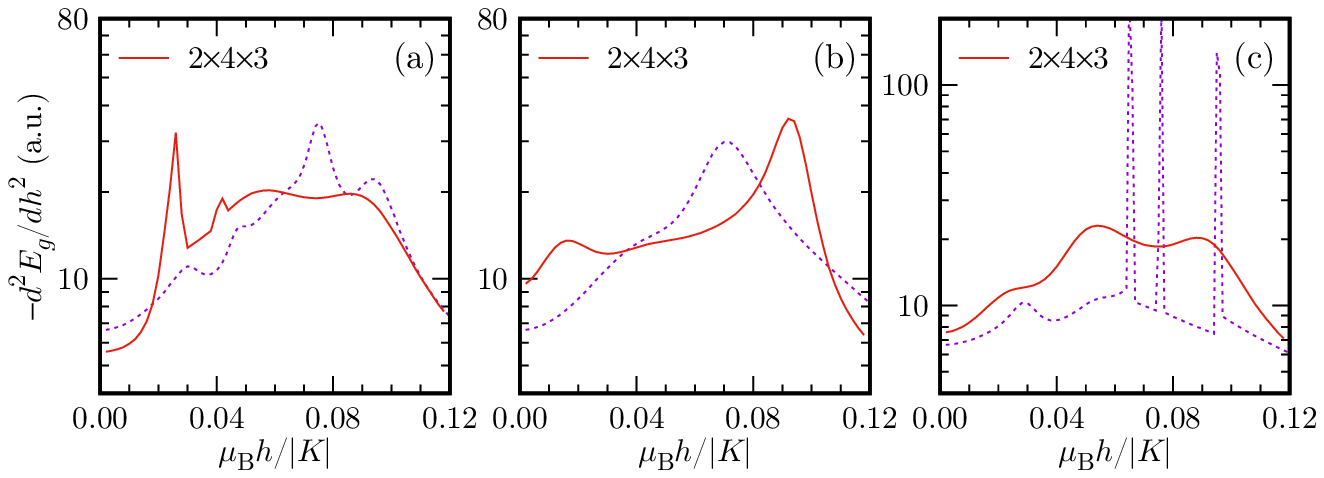}
\caption {
 Second derivative of the ground state energy ($d^2E_g/dh^2$) 
 with respect to the magnetic field $h$ 
 along (a) the $a$ axis, (b) the $b$ axis, and (c) the $c$ axis 
 for the periodic $2 \times 4 \times 3$ cluster.  
 Schematic diagram of the periodic $2\times L_1 \times L_2$ cluster is shown
 in Fig.~\ref{fig_dmrg}(a). 
 For comparison, the results for the $C_3$ rotationally symmetric 24-site cluster 
 is also shown by dotted lines. 
 All results are obtained for the $K$-$J_3$ model with $K<0$
 and $J_3/|K|=0.1$ calculated using the ED method.
 }
\label{fig_dc}
\end{figure}

To check the cluster geometry dependence on the magnetic phase transition
of the $K$-$J_3$ model in the magnetic field,
here we perform the ED calculation on a periodic $2 \times 4 \times 3$ cluster 
[for the geometry of the cluster, see Fig.~\ref{fig_dmrg}(a)].
Unlike the 24-site cluster shown in Fig.~\ref{fig_hc}(b), 
the $2 \times 4 \times 3$ cluster does not have the $C_3$ rotation symmetry.
Instead, it is invariant under the $C_2$ rotation.
Figure~\ref{fig_dc} shows the results of $-d^2E_g/dh^2$ 
in the $a$-, $b$-, and $c$-axis fields for $J_3/|K|=0.1$, 
which are also compared with those for the $C_3$ rotationally symmetric 24-site cluster. 
Because these clusters are still too small to avoid the finite size effect, 
the field dependence of $-d^2E_g/dh^2$ evidently depends on the cluster geometry.
Nevertheless, the overall shapes are consistent with each other except 
that there are two peaks in the $2 \times 4 \times 3$ cluster, while there are 
three peaks in the $C_3$ rotationally symmetric 24-site cluster, 
in the large $c$-axis field $\mu_B h/|K| > 0.05$. 
The emergence of the intermediate phase (IP) in the $a$- and $c$-axis fields, 
and the absence of any IP in the $b$-axis field are identified 
in both clusters.
However, the number of consecutive IPs is hardly determined.
It depends strongly on the geometry and size of clusters.

\section{DMRG calculation}\label{Sec:DMRG}

\begin{figure}[h!]
\centering
\includegraphics[width=.7\columnwidth]{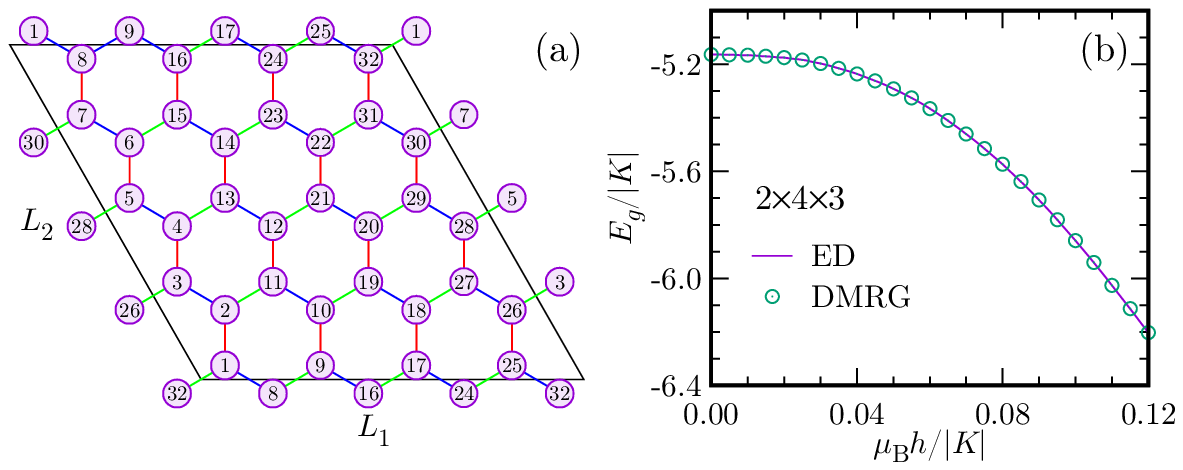}
\caption {
 (a) Schematic diagram of a periodic $2\times L_1 \times L_2$ cluster 
  with $L_1=L_2=4$. 
 (b) Ground state energy $E_g$ as a function of the magnetic field $h$ along the $a$ axis 
 for the $K$-$J_3$ model with the ferromagnetic Kitaev interaction ($K<0$) 
 and $J_3/|K|=0.1$ calculated on a periodic $2\times 4 \times 3$ cluster
   using the ED and DMRG methods.
}
\label{fig_dmrg}
\end{figure}

To solve the $K$-$J_3$ model with the density matrix renormalization group (DMRG) method,
we consider a periodic $2 \times 4 \times 4$ cluster shown in Fig.~\ref{fig_dmrg}(a).
To verify the relevance of the DMRG calculation in the two-dimensional $K$-$J_3$ model,
we also perform the DMRG calculation of a periodic $2 \times 4 \times 3$ cluster and 
compare it with the ED calculation.
Figure~\ref{fig_dmrg}(b) shows the energies calculated by the ED and DMRG methods 
for various values of the $a$-axis field when we keep up to $m=2000$ eigenstates with largest eigenvalues of the reduced 
density matrix of the ground state in the DMRG calculation.
We check that the energy is converged with the accuracy less than $10^{-7}|K|$.
Despite of the two-dimensional system, 
the DMRG calculation is thus adequate to obtain the ground state of the $K$-$J_3$ model.
In the case of the periodic $2 \times 4 \times 4$ cluster, we increase the number of 
density-matrix eigenstates kept up to $m=2500$ for better convergence. 
Because of the large computational cost, we perform the calculation with a 
mild truncation error of the ground state energy around $10^{-7}|K| \sim 10^{-4}|K|$.

\section{Classical Monte-Carlo Calculation}

\begin{figure}[h!]
\centering
\includegraphics[width=.7\columnwidth]{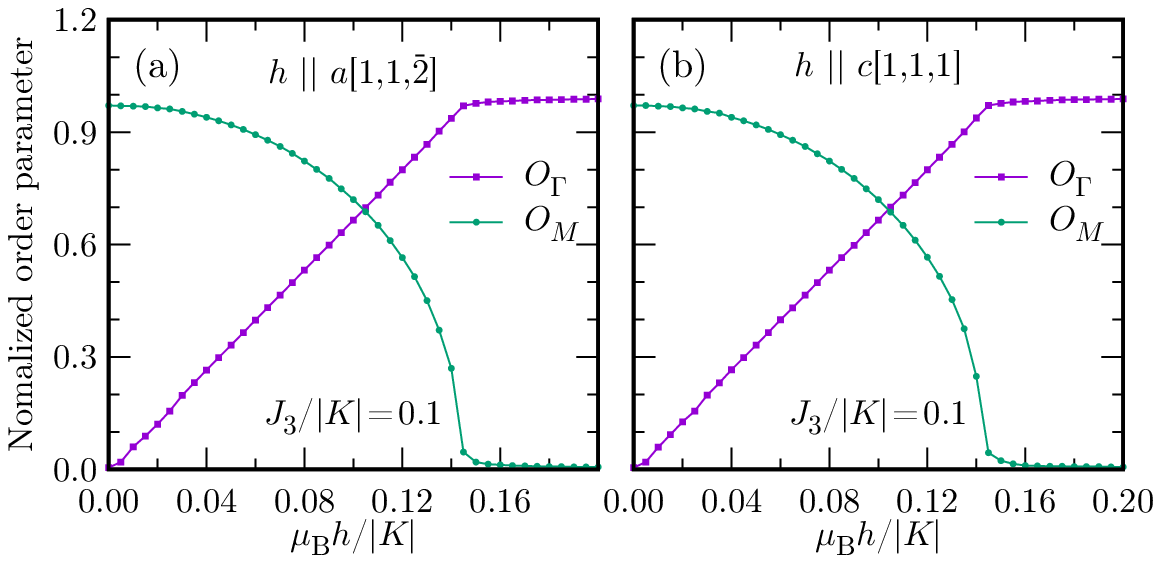}
\caption {
  Evolution of order parameters for the zigzag order phase ($O_{M}$) and 
  the polarized phase ($O_{\Gamma}$) 
  as a function of the external magnetic field along (a) the $a$ axis and (b) the $c$ axis 
  for the classical $K$-$J_3$ model with $J_3/|K|=0.1$. 
  The temperature is set to be $0.01 |K|/k_B$ and a periodic $2\times24\times24$ cluster is used in the classical MC calculation.
}
\label{fig_mc}
\end{figure}

To understand the magnetic phase transition of the classical $K$-$J_3$ model
in the presence of the magnetic field,
we consider a periodic $2\times 24 \times 24$ cluster 
[for the geometry of the cluster, see Fig.~\ref{fig_dmrg}(a)] and
perform the classical Monte-Carlo (MC) calculation with the standard Metropolis algorithm.
Figure~\ref{fig_mc} shows the evolution of order parameters $O_\Gamma$ and $O_M$ 
(i.e. static SC function $\left< \mathbf{S}_\mathbf{-q}\cdot\mathbf{S}_\mathbf{q}\right>$ at the $\Gamma$ and 
$M$ points, respectively) for
the polarized phase and the zigzag order phase for $J_3/|K|=0.1$
at temperature $k_BT/|K|=0.01$ in the $a$- and $c$-axis fields
calculated with 40000 MC steps after 20000 MC steps for thermalization.
As the magnetic field increases, 
the order parameter for the zigzag order phase decreases with increasing 
the slop until losing all intensity,
while that for the polarized phase almost linearly increases until it is completely saturated. 
No indication of an intermediate phase is observed 
during the phase transition from the zigzag order phase to the polarized phase.

\section{Static spin correlation, Magnetization,
and dynamical spin correlation}

To gain an insight on the IP, 
we calculate the static SC functions $\left< \mathbf{S}_i \cdot \mathbf{S}_j\right>$,
magnetizations, and dynamical SC functions for the $K$-$J_3$ model 
on the periodic 24-site cluster shown in Fig.~\ref{fig_hc}(a)
in the presence of the magnetic field. 

\begin{figure}[h!]
\centering
\includegraphics[width=.9\columnwidth]{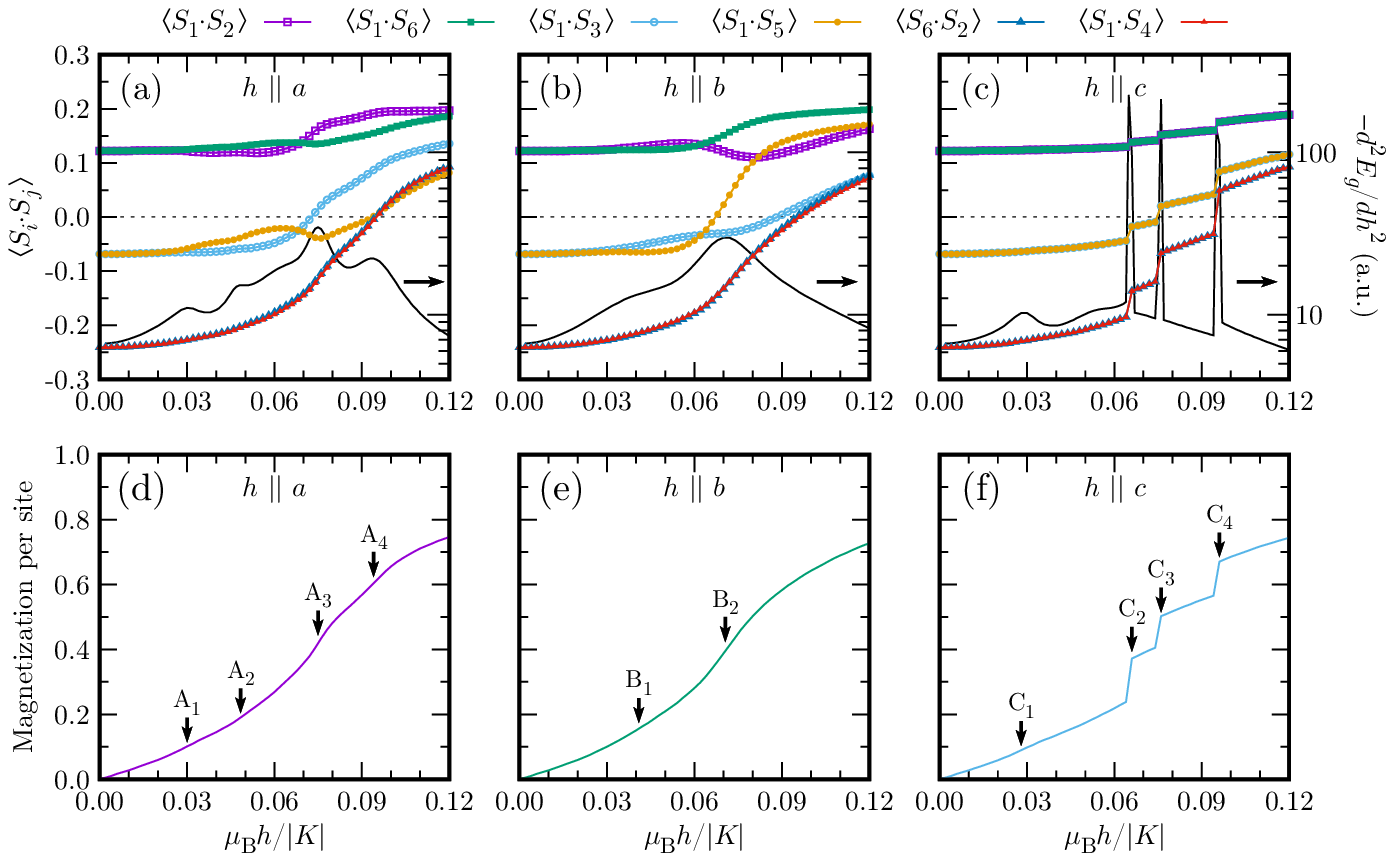}
\caption {
 (a)--(c) Spin correlation functions $\left< \mathbf{S}_i\cdot\mathbf{S}_j\right>$
 between NN spins 
 ($\left< \mathbf{S}_1 \cdot \mathbf{S}_2\right>$,  
 $\left< \mathbf{S}_1 \cdot \mathbf{S}_6\right>$),
 second NN spins 
 ($\left< \mathbf{S}_1 \cdot \mathbf{S}_3\right>$,
 $\left< \mathbf{S}_1 \cdot \mathbf{S}_5\right>$),
 and third NN spins 
 ($\left< \mathbf{S}_6 \cdot \mathbf{S}_3\right>$,
 $\left< \mathbf{S}_1 \cdot \mathbf{S}_4\right>$)
 as a function of the magnetic field $h$ along 
 (a) the $a$ axis, (b) the $b$ axis, and (c) the $c$ axis. 
 The locus of spins in the cluster are indicated in Fig.~\ref{fig_hc}(a).
 (d)--(f) Magnetization per site along the field direction 
 as a function of the magnetic field $h$ along
 (d) the $a$ axis, (e) the $b$ axis, and (f) the $c$ axis. 
 For comparison, second derivative of the ground state energy ($d^2E_g/dh^2$) with respect to the field 
 is also shown in (a)--(c). 
 Arrows in (d)--(f) indicate the positions of peaks A$_{1-4}$, B$_{1.2}$, 
 and C$_{1-4}$ of $-d^2E_g/dh^2$ [also see Figs.~1(b) and 1(c) in the main text]. 
 All results are obtained for the $K$-$J_3$ model with the ferromagnetic Kitaev interaction ($K<0$) 
 and $J_3/|K|=0.1$ calculated 
 on a periodic 24-site cluster using the ED method. 
}
\label{fig_ss}
\end{figure}

Figures~\ref{fig_ss}(a)--\ref{fig_ss}(c) show the static SC functions in the $a$-, $b$-, and $c$-axis fields. 
In the case of $J_3/|K|=0.1$ with no magnetic field where the zigzag order is dominant, 
strong FM Kitaev and AFM $J_3$ interactions 
lead to positive SC function ($\approx0.12$) among nearest neighboring (NN) spins and 
negative SC function ($\approx-0.24$) among third NN spins. 
On the other hand, the SC function among second NN spins is as large as $-0.068$.
This is because two spins at second NN sites can simultaneously 
belong to only one of the three FM zigzag chains determined by the ordering
momenta, $M_1$, $M_2$, and $M_3$ points. Thus, 
the SC function among second NN spins is the average value of two AFM and one FM correlations.

When the magnetic field is weak enough ($\mu_Bh/|K| \lessapprox 0.06$), 
the zigzag order is still dominant and the static SC functions are almost robust  
[see Figs.~\ref{fig_ss}(a)--\ref{fig_ss}(c)].
In the $c$-axis field, two consecutive IPs take place with increasing the field.
As shown in Fig.~\ref{fig_ss}(c), all SC functions vary discontinuously at the boundaries of the IPs. 
With further increasing the field, the SC functions eventually become all positive 
because all spins are almost polarized along the field direction.
In the $a$-axis field, the discontinuous change of the SC functions is somewhat suppressed but
the steep variation of the SC functions still appears around the IP boundaries 
and their slope becomes slightly moderate in the IP region [see Fig.~\ref{fig_ss}(a)]. 
In contrast, in the $b$-axis field, the SC functions change abruptly around the critical field and 
then increase continuously, as shown in Fig.~\ref{fig_ss}(b).

Figures~\ref{fig_ss}(d)--\ref{fig_ss}(f) show the magnetization per site along the respective field direction 
as a function of the magnetic field along the $a$, $b$, and $c$ axes. 
In the $c$-axis field, the step-like magnetization is manifested around the IPs. 
This is reminiscence of the magnetization plateau frequently observed 
in frustrated quantum magnets.
When the field is applied away from the $c$ axis, the step-like feature 
becomes diminished. 
However, we can still notice that the slope of the magnetization curve changes somewhat 
discontinuously around the boundaries of the IP
in the $a$-axis field, 
while the magnetization simply increases smoothly in the $b$-axis field.

\begin{figure}[h!]
\centering
\includegraphics[width=.9\columnwidth]{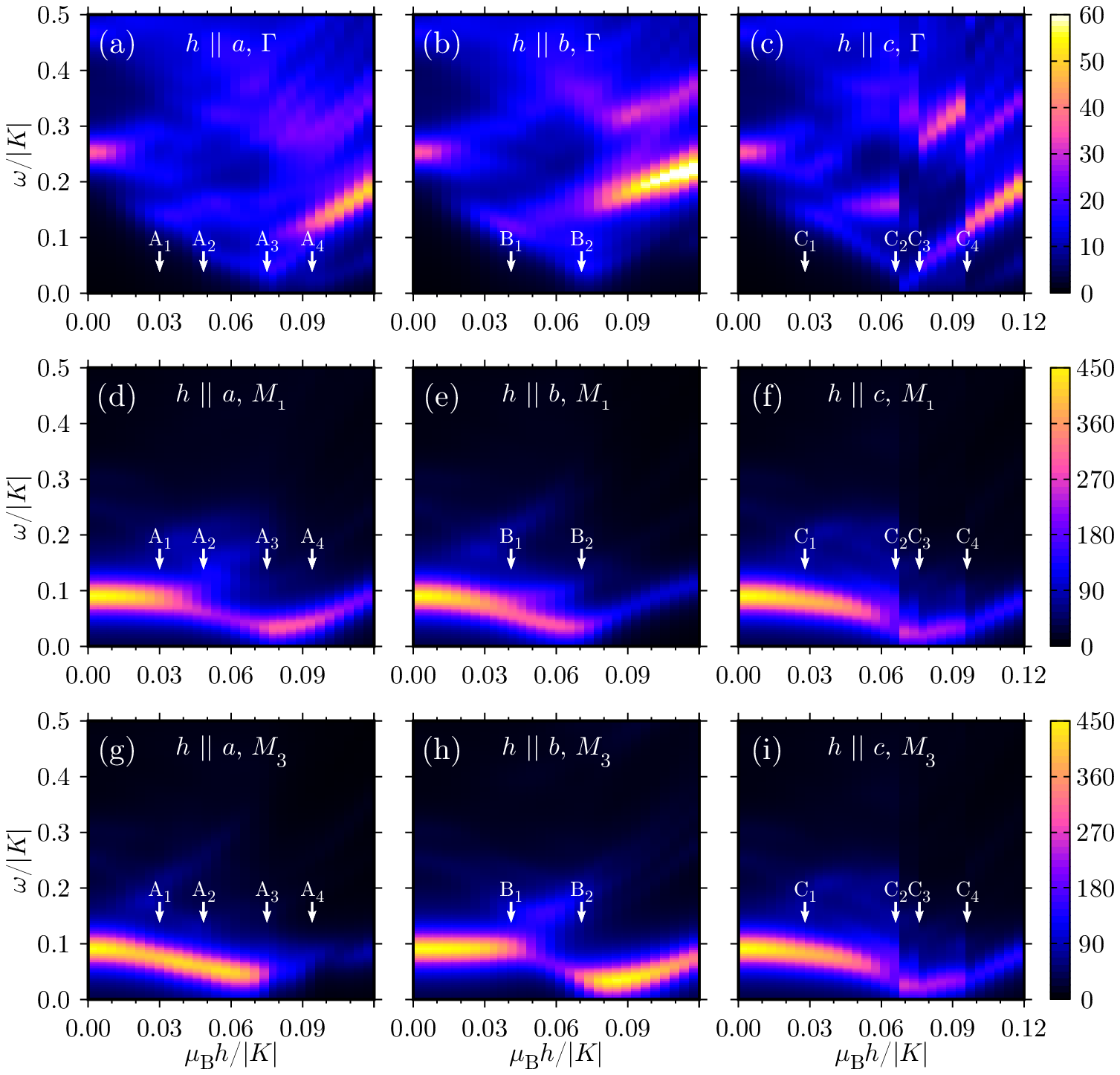}
\caption {
 Dynamical spin correlation functions $C_{\mathbf{q}}(\omega)$
 as a function of the magnetic field $h$ along the $a$ axis
 at (a) the $\Gamma$ point, (d) the $M_1$ point, and (g) the $M_3$ point, 
 along the $b$ axis at (b) the $\Gamma$ point, 
 (e) the $M_1$ point, and (h) the $M_3$ point, 
 and along the $c$ axis at (c) the $\Gamma$ point, (f) the $M_1$ point, and (i) the $M_3$ point. 
 White arrows indicate the positions of peaks A$_{1-4}$, B$_{1,2}$, and 
 C$_{1-4}$ of $-d^2E_g/dh^2$ [see Fig.~1(b) and 1(c) in the main text].
 For clarity, the elastic contribution of $C_{\mathbf{q}}(\omega)$ at $\omega=0$ is omitted. 
 All results are obtained for the $K$-$J_3$ model with the ferromagnetic Kitaev interaction ($K<0$) 
 and $J_3/|K|=0.1$ calculated 
 on a periodic 24-site cluster using the ED method. 
}
\label{fig_sw}
\end{figure}

Dynamical SC function $C_{\mathbf{q}}(\omega)$ at momentum $\mathbf{q}$
and energy $\omega$ is given as
\begin{equation}
C_{\mathbf{q}}(\omega)=
-\frac{1}{\pi}\textrm{Im} \left[ 
\sum_{a \in \{x,y,z\}}
\left< 
S_{\mathbf{-q},a}\frac{1}{\omega - H + E_g + i\delta}S_{\mathbf{q},a} 
\right>
\right],
\end{equation}
where $E_g$ is the ground state energy and $\delta\, (=0.025|K|)$ is the broadening parameter. 
Figure~\ref{fig_sw} shows $C_{\Gamma}(\omega)$, $C_{M_1}(\omega)$,
and $C_{M_3}(\omega)$ for the $K$-$J_3$ model in the $a$-, $b$-, and $c$-axis fields.
Note that $C_{M_2}(\omega)$ is exactly the same as $C_{M_1}(\omega)$ 
in these three magnetic fields due to the symmetry.

As shown in Figs.~\ref{fig_sw}(a)--\ref{fig_sw}(c), 
$C_{\Gamma}(\omega)$ exhibits strong spectral weight at $\omega/|K|\approx 0.25$ when $h=0$, 
determining the excitation gap in the absence of the field. 
Applying the magnetic field, the spectral weight is spread over a wide region of $\omega$
in any field direction.
The excitation gap is decreased with increasing the field until the zigzag order phase is no longer robust. 
The excitation gaps are about $0.038|K|$, $0.045|K|$, and $0.046|K|$ 
at the phase boundaries corresponding to the A$_3$ position ($\mu_Bh/|K|\approx 0.075$),
B$_2$ position ($\mu_Bh/|K|\approx 0.071$), and C$_2$ position ($\mu_Bh/|K|\approx 0.066$)
in the $a$-, $b$-, and $c$-axis fields, respectively. 
In the $b$-axis field, the excitation gap begins to increase with further increasing 
the field above the B$_2$ position.
In the $a$-axis field, the excitation gap similarly increases 
after the A$_3$ position except that a weak spectral weight arises 
below the main excitation spectra [see Fig.~\ref{fig_sw}(a)]. 
In the $c$-axis field, the excitation spectra change discontinuously 
around the phase boundaries, similar to the evolution of the static SC function and the magnetization shown in 
Figs.~\ref{fig_ss}(c) and \ref{fig_ss}(f), respectively.
The excitation gap suddenly drops from $0.046|K|$ to zero at the C$_2$ position
and then simply increases with further increasing the field. 
Notice also that as in the case of the $a$-axis field, a weak spectral weight arises 
below the main spectral wight.
The excitation gap detected from this emergent weak spectral weight closes at the phase boundaries determined by 
the C$_3$ position ($\mu_Bh/|K|\approx 0.076$) and C$_4$ position ($\mu_Bh/|K|\approx 0.096$), 
and increases monotonically in the polarized phase.

As shown in Figs.~\ref{fig_sw}(d)--\ref{fig_sw}(i), $C_{M_1}(\omega)$ and $C_{M_3}(\omega)$ show 
strong spectral weight at $\omega/|K|\approx 0.09$ in the absence of the field.
In contrast with $C_{\Gamma}(\omega)$, these spectral weights 
are hardly spread even in the finite fields when the zigzag order is stabilized. 
In the $a$-axis field, $C_{M_1}(\omega)$ 
has a minimum excitation gap ($\approx 0.025|K|$) around the phase boundary between the zigzag order phase and the IP. 
In the $b$-axis field, $C_{M_3}(\omega)$
has a minimum excitation gap ($\approx 0.033|K|$) at $\mu_Bh/|K|\approx 0.08$, 
which is slightly larger than the critical field $\mu_Bh/|K|\approx 0.071$
(B$_2$ position).
In the $c$-axis field, 
the minimum excitation gap ($\approx 0.022|K|$) of $C_{M_1}(\omega)$ and $C_{M_3}(\omega)$ 
appears at $\mu_Bh/|K|\approx 0.076$, 
the phase boundary between the two IPs (C$_3$ position). 
Inside these two IPs, the variation of the excitation gap is not much strong.

\section{Entanglement Entropy behaviors}

\begin{figure}[h!]
\centering
\includegraphics[width=.7\columnwidth]{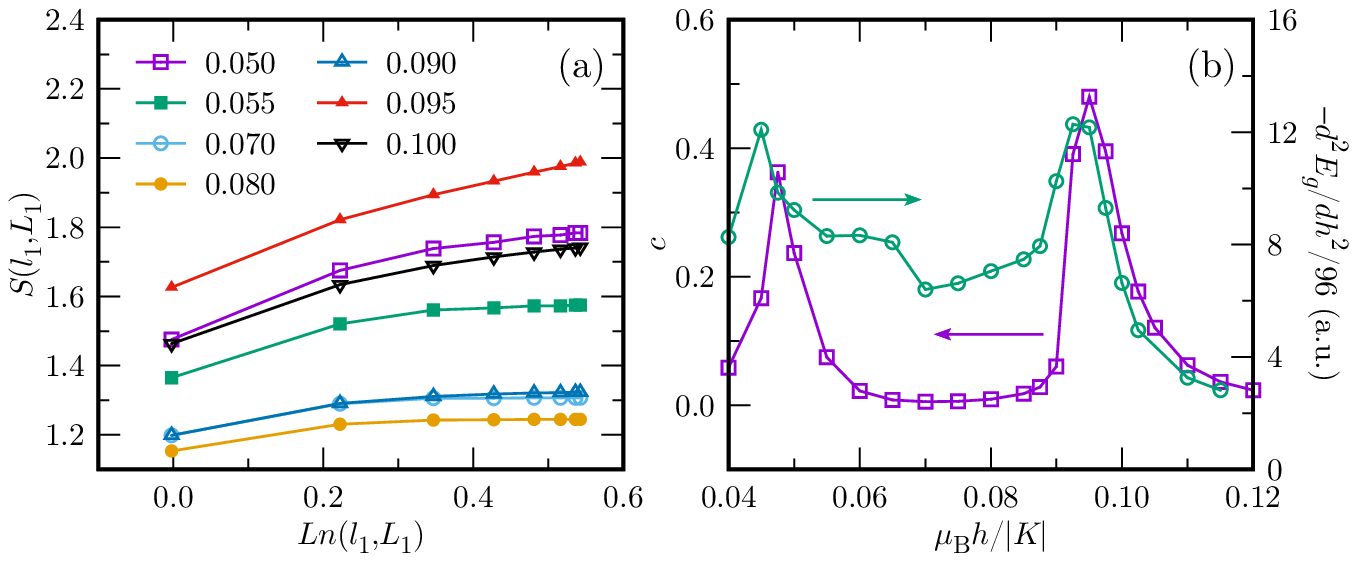}
\caption {
(a) The Entanglement entropy (EE) 
as a function of the subsystem length $l_1$ along the $L_1$ direction 
for various $c$-axis field strengths ($\mu_B h/|K|=0.05$, $0.055$, $0.07$,
$0.08$, $0.09$, $0.095$, and $0.1$).
$Ln(l_1,L_1)$ in the horizontal axis is defined as $Ln(l_1,L_1) = \frac{1}{3} \ln 
\left[ \frac{L_1}{\pi}\sin\left( \frac{\pi l_1}{L_1}\right) \right]$.
For $l_1>2$, the EEs are fitted reasonably well by 
the linear function of $Ln(l_1,L_1)$.
(b) Second derivative of the ground state energy per site ($d^2E_g/dh^2/N$)
with respect to the $c$-axis field $h$ and 
the central charge $c$ estimated by fitting the EEs with Eq.~(\ref{eq_EE}).
All results are obtained for the $K$-$J_3$ model with $K<0$ and $J_3/|K|=0.1$
on a periodic $ 2\times L_1 \times L_2$ cluster with $L_1 = 16$ and $L_2 = 3$ 
(thus the system size being $N=96$)
calculated using the DMRG method.
}
\label{fig_ee}
\end{figure}

To further investigate the characteristic feature of the IP, 
we explore the entanglement entropy (EE) of the $K$-$J_3$ model on 
a periodic $2\times L_1 \times L_2$ cluster with $L_1=16$ and $L_2=3$ by 
the DMRG method keeping $m=1500$ (see Sec.~\ref{Sec:DMRG}).
We calculate the von Neumann entanglement entropy of the subsystem by varying 
the subsystem length $l_1$ from $l_1=1$ to $L_1/2$ along the $L_1$ direction, i.e., 
\begin{equation}
S(l_1,L_1) = -\textrm{Tr}_{l_1} \rho_{l_1} \ln \rho_{l_1},
\end{equation}
where $\textrm{Tr}_{l_1}$ is the trace over all basis on the subsystem 
and $\rho_{l_1}$ is the reduced density matrix of the subsystem.
Figure~\ref{fig_ee}(a) shows $S(l_1,L_1)$ for various $c$-axis 
field strengths ($\mu_B h/|K|=0.05$, $0.055$, $0.07$, $0.08$, $0.09$, $0.95$,
and $0.1$) when $J_3/|K|=0.1$.
The calculated EEs for $l_1>2$ can be fitted reasonably well with  
the prediction of the conformal field theory (CFT) for
a $1+1$ dimensional critical system, i.e., 
\begin{equation}
\label{eq_EE}
S(l_1,L_1) = \frac{c}{3} \ln 
\left[ \frac{L_1}{\pi}\sin\left( \frac{\pi l_1}{L_1}\right) \right] +s',
\end{equation}
where $c$ is the central charge of the CFT 
and $s'$ is a nonuniversal constant~\cite{Nishimoto2011sm,Jiang2013sm}.
As shown in Fig.~\ref{fig_ee}(b), $-d^2E_g/dh^2$ shows two relatively sharp peaks 
at $\mu_Bh/|K|\approx0.045$ and $0.095$, and one broaden peak at $\mu_Bh/|K|\approx0.06$,  
thus supporting that there exist one or two IPs emerging 
also in the periodic $2\times 16 \times 3$ cluster. 
Interestingly, the obtained $c$ value shown in Fig.~\ref{fig_ee}(b) is almost zero 
for $0.06\alt \mu_B h/|K| \alt 0.09$ in the IP.
This infers that the IP is gapped, 
which  is in contrast with the $K$-$J$ model with antiferromagnetic $K$ and 
ferromagnetic $J$ model where the possible IP is proposed to be 
the gapless $U(1)$ spin liquid with $c=2$~\cite{Jiang2019sm}.
In addition, we find that $c$ has a finite value at the magnetic field around which 
$-d^2E_g/dh^2$ exhibits the relatively sharp peaks. 
This implies that the excitation gap is closed at the critical fields of
the phase transition.

\end{document}